\documentclass[nonacm,sigconf]{acmart}

\AtBeginDocument{%
	\providecommand\BibTeX{{%
			\normalfont B\kern-0.5em{\scshape i\kern-0.25em b}\kern-0.8em\TeX}}}

\usepackage{hyperref}
\usepackage{tikz}
\usepackage{soul}
\usepackage[nohyperlinks, nolist]{acronym}
\usepackage{tabularx}
\usepackage{enumitem}
\usepackage{amsmath}
\usepackage{xspace}
\usepackage{calc}
\usepackage{algorithm}
\usepackage[noend]{algpseudocode}
\usepackage{dcolumn}
\usepackage{cleveref}
\usepackage{etoolbox}
\usepackage{pifont}

\newcolumntype{d}[1]{D{.}{.}{#1}}

\renewcommand\footnotetextcopyrightpermission[1]{} 

\DeclareMathAlphabet{\mathpzc}{T1}{pzc}{m}{it}
\newcommand{\treename}{PITS\xspace}

\newcommand{\adv}{$\mathcal{A}$\xspace}
\newcommand{\monitored}{$\mathcal{M}$\xspace}
\newcommand{\interacter}{$\mathcal{I}$\xspace}
\newcommand{\notary}{$\mathcal{N}$\xspace}
\newcommand{\validator}{$\mathcal{V}$\xspace}
\newcommand{\epoch}{$ep$\xspace}
\newcommand{\subepoch}{$subep$\xspace}
\newcommand{\depthts}{$size_{ts}$\xspace}
\newcommand{\depthparity}{$depth_p$\xspace}
\newcommand{\depthupdate}{$depth_u$\xspace}
\newcommand{\sizeparity}{$size_p$\xspace}
\newcommand{\elog}{$l$\xspace}
\newcommand{\eloghash}{$h(l)$\xspace}
\newcommand{\elogsepoch}{$L_{ep}$\xspace}

\newcommand{\elogts}{$l_{ts}$\xspace}
\newcommand{\elogreceipt}{$l_{rec}$\xspace}
\newcommand{\tree}{$t$\xspace}
\newcommand{\treeepoch}{$t_{ep}$\xspace}
\newcommand{\treeroot}{$t_r$\xspace}
\newcommand{\treeparity}{$t_p$\xspace}
\newcommand{\treeupdate}{$t_u$\xspace}
\newcommand{\treesecret}{$t_{sk}$\xspace}

\begin{acronym}
	\acro{poi}[PoI]{Proof of Inclusion}
	\acroplural{poi}[PoIs]{Proofs of Inclusion}
	\acro{mht}[MHT]{Merkle Hash Tree}
	\acro{smt}[SMT]{Sparse Merkle Tree}
\end{acronym}

\begin{document}

\title[Accountability of Things: Large-Scale Tamper-Evident Logging for Smart Devices]{Accountability of Things:\\Large-Scale Tamper-Evident Logging for Smart Devices}

\author{David Koisser}
\affiliation{%
	\institution{Technical University of Darmstadt}
	\city{Darmstadt}
	\country{Germany}
}
\email{david.koisser@trust.tu-darmstadt.de}

\author{Ahmad-Reza Sadeghi}
\affiliation{%
	\institution{Technical University of Darmstadt}
	\city{Darmstadt}
	\country{Germany}
}
\email{ahmad.sadeghi@trust.tu-darmstadt.de}

\keywords{Tamper-Evident Logging, Audit and Accountability, Digital Forensics}


\begin{abstract}
Our modern world relies on a growing number of interconnected and interacting devices, leading to a plethora of logs establishing audit trails for all kinds of events.
Simultaneously, logs become increasingly important for forensic investigations, and thus, an adversary will aim to alter logs to avoid culpability, e.g., by compromising devices that generate and store logs.
Thus, it is essential to ensure that no one can tamper with any logs without going undetected.
However, existing approaches to establish tamper evidence of logs do not scale and cannot protect the increasingly large number of devices found today, as they impose large storage or network overheads.
Additionally, most schemes do not provide an efficient mechanism to prove that individual events have been logged to establish accountability when different devices interact.

This paper introduces a novel scheme for practical large-scale tamper-evident logging with the help of a trusted third party.
To achieve this, we present a new binary hash tree construction designed around timestamps to achieve constant storage overhead with a configured temporal resolution.
Additionally, our design enables the efficient construction of shareable proofs, proving that an event was indeed logged.
Our evaluation shows that---using practical parameters---our scheme can localize any tampering of logs with a sub-second resolution, with a constant overhead of $\sim$8KB per hour per device.
\end{abstract}

\maketitle

\section{Introduction}
\label{sec:intro}
As our world is getting increasingly digital and interconnected, cyber-physical systems and the Internet of Things (IoT) are becoming prevalent.
Indeed, most aspects of our lives these days involve interactions with on-line services and smart devices equipped with sensors collecting information as well as autonomously interacting with each other.
These devices generate a large number of logs that allow analyzing events retrospectively, which is becoming increasingly important for forensic investigations~\cite{cpsforensicssurvey,iotforensicsarticle,iotforensicssurvey}, ranging from simple causal analysis, e.g., to identify faulty devices and ascertain culpability in a legal incident.

Logs play an important role in many different settings: 
In environments with IoT devices for physical security like smart locks, smart card readers for access control, and nearby devices logging correlated events~\cite{iotforensicsarticle,iotforensicssurvey}, log files of devices can be used for forensic analysis to reveal possible culprits in case of security incidents.
In machine-to-machine interactions in an industrial Internet~\cite{i4forensics,cpsforensicssurvey}, machines from different manufacturers and operators may sense the properties of a product and interact with passing on products for different manufacturing stages.
However, in case of a faulty product, the logs of the involved machines can reveal where the manufacturing went wrong to establish culpability.
Other examples include smart grid~\cite{smartgridforensics}, smart traffic~\cite{smarttrafficforensics}, and smart healthcare devices~\cite{healthcareforensics}.
Clearly, the logs of such devices are crucial for forensic investigations in the digital age.

While logs are crucial for causal analysis, they also become an attractive target for adversaries.
Logs can be altered to evade any culpability in an incident, e.g., by an insider authorized to access the respective devices or via  compromise.
The adversary could then modify, delete, or insert misleading logs, leading any forensic investigation astray.
Thus, it is important to ensure any tampering with the logs is evident.
This is challenging, as very fine-granular integrity proofs do not scale well, such as integrity proofs on a per-log basis.
Conversely, very coarse integrity proofs, e.g., over a week's worth of logs, make it difficult to locate the tampering meaningfully, e.g., anytime in the respective week.
Further, when systems of devices autonomously interact with each other, it is important to establish \emph{accountability} between them, i.e., each device can establish proof of their interactions with each other.
For example, suppose the log of an interaction was deleted on a compromised device.
In that case, it is highly advantageous for the other device to be able to present proof that the interaction was originally logged on the compromised device, i.e., provide a \emph{log receipt}.

There is a variety of proposals to address this challenge, yet, they have crucial shortcomings.
A simple solution coming to mind is simply sending all logs to a trusted server for storage and later retrieval for forensic analysis~\cite{rwmRemote2,rwmRemote3}.
However, this requires a constant and stable connection to the trusted server at all times to ensure security, which is hard to guarantee, especially if the adversary is able to manipulate the device. 
Another simple solution is to involve a blockchain~\cite{sutton2017blockchain,cucurull2016distributed,ahmad2018towards,ahmad2019blocktrail,aniello2017prototype,pourmajidi2018logchain}. 
However, permissionless blockchains (e.g., Bitcoin) are notoriously slow and expensive, while the use of permissioned blockchains multiplies the storage and network overheads of remote logging services; thus, these approaches are unfit to cover the logs generated by a plethora of devices with meaningful accuracy.
Another approach is to use specialized hardware for logging.
Nevertheless, the settings described above may consist of heterogeneous devices, and the use of auxiliary hardware~\cite{HWworm,HWtpm,HWhardlog} or specific hardware extensions may not be feasible~\cite{HWsgx,HWcustos} for all devices.
Previous works on tamper-evident data structures use symmetric cryptography and hash functions~\cite{bellare1997forward,schneier1998cryptographic,schneier1999secure}.
Yet, these schemes induce large storage overheads and cannot efficiently generate log receipts.
Other works leverage forward-secure aggregated signatures~\cite{ma2009new,yavuz2009baf,yavuz2012efficient} for efficient tamper-detection of logs.
Yet, they do not support efficient log receipts or require large storage overheads.
Hash tree structures have been used to generate log receipts that are gossiped throughout the network~\cite{crosby2009efficient,pulls2015balloon}. 
However, this requires the monitored devices to store an ever-growing hash tree, induces a lot of network overhead, and depends on one entity seeing all log receipts to establish the holistic integrity of the monitored device. 
We will discuss the related work in detail in \Cref{sec:rw}. 
Nevertheless, providing large-scale tamper-evident logging for smart devices, including the possibility of generating log receipts, remains an open problem.

We present a novel scheme for practical large-scale tamper-evident logging for smart devices with the help of a trusted third-party. 
To achieve this, we construct a novel binary hash tree structure designed around log timestamps. 
This tree stores additional information that enables checking the integrity of a set of logs at a chosen time resolution, e.g., sub-second, while the size of the stored integrity information stays constant, regardless of the number of logs. 
This allows localizing any tampering in a set of logs within the specified resolution, e.g., the specific second the logs were altered.
Further, our scheme allows the efficient construction of log receipts.
\\~\\
\noindent Our main contributions include:

\begin{itemize}
	\item We design a novel hash tree, which is able to store integrity information on a specified time resolution efficiently.
	Our tree requires constant storage overhead on a trusted log service and allows the construction of efficient log receipts.
	\item We define our logging scheme with interactions between the different actors, which further provides \emph{forward integrity}, i.e., detecting tampering of unsent logs even after the device was compromised, and an efficient way to distribute log receipts to prevent impractical waiting times.
	\item We implement a prototype of our approach and show that our scheme needs a constant $\sim$8KB per hour and per device---using practical parameters---to store sub-second resolution tamper-evident integrity information.
	We also show that a single server is able to process around $700\,000$ logs per second, and thus demonstrate the practicality of our scheme.
\end{itemize}


\section{System Model}
\label{sec:model}
There are four different actors in our scheme.
One actor is the monitored node \monitored, which will log events that are to be protected against tampering.
Events can refer either to \emph{internal} events, something \monitored has observed, or an event involving an interaction with another entity, i.e., the interacting node \interacter.
The log notary \notary is trusted and will protect the integrity of \monitored's logs to prevent any later tampering.
For example, \notary could be the manufacturer of device \monitored or a regulatory authority.
We assume that the connection between \monitored and \notary is unreliable and that they may communicate infrequently.
If logs describe interactions, \interacter as the other entity having a direct exchange with \monitored will request proofs of logs corresponding to these exchanges from \notary.
These proofs allow \interacter to prove that an event has happened, e.g., to hold \monitored accountable in a later dispute.
Nevertheless, when there is a bilateral interaction between two nodes, they could both act as a monitored and interacting node, mutually collecting proofs of their bilateral interaction.
The validator \validator aims to validate the integrity of logs on \monitored and prove any tampering.
We assume that communication channels between entities are secured (e.g., via TLS).
Finally, we assume there is a coarse time synchronization between all entities, i.e., all clocks have differences in the order of seconds to roughly agree on time epochs.

\subsection{Adversary Model}
\label{sec:advmodel}

The goal of adversary \adv is to tamper with the logs on \monitored by modifying, deleting, or inserting logs.
\adv wants to stay undetected to cover up some prior event that was logged, e.g., an authorized access that would incriminate \adv.
Thus, \adv either is an insider authorized to access \monitored or has the ability to compromise \monitored.
For typical assumptions, we assume \adv cannot forge digital signatures, cannot find collisions in cryptographic hash functions or Merkle tree constructions, and cannot break the secure channels between entities from the outside.

\subsection{Requirements}
\label{sec:reqs}
To formalize the settings outlined in the Introduction, we aim to design a tamper-evident logging scheme with the following requirements:
\begin{enumerate}
	\item[R.1] \label{req1:tamper}\emph{Tamper-detection with chosen time resolution:} 
	The scheme shall allow detection of any tampering with logs.
	Different use cases may require detecting any tampering with different time granularity; thus, the time resolution of the scheme shall be set as a parameter.
	This allows to establish a trade-off between time resolution and induced overheads.
	Note that setting the time resolution too low (e.g., one millisecond) is impractical, as this would require high-accuracy clock synchronization between devices.
	\item[R.2] \label{req2:receipts}\emph{Efficient log receipts:} 
	Devices shall be able to establish accountability between each other with regard to their interactions.
	Thus, the scheme shall enable efficient generation and storage of log receipts.
	This allows one device to keep proof that another was participating in an event.
	\item[R.3] \label{req3:overhead}\emph{Low overheads on logging device:} 
	As devices might be heterogeneous in terms of hardware, the scheme shall induce insignificant overheads on the logging devices.
	Thus, the primary overhead shall be on the side of the log notary.
\end{enumerate}


\section{Design}
\acresetall
\label{sec:desgin}
To describe our scheme, we first give an overview of the entire system, including a high-level description of our novel integrity tree construction and how the actors interact with each other.
In \Cref{sec:tree}, we will describe our integrity tree construction in detail, and finally, in \Cref{sec:ops}, we will thoroughly define the operations executed between the actors.

\subsection{Overview}
At the core of our scheme is a novel binary hash tree structure similar to the Merkle hash tree~\cite{mht}.
Like Merkle trees, this structure is a cryptographic accumulator, i.e., a way to represent a big set of data elements in a compact way.
While the individual contained elements cannot be retrieved, an accumulator allows for the construction of \acp{poi}, which prove an individual element is part of the set represented by, e.g., a single hash in binary hash trees, called the tree's root.
Our tree's construction is designed around timestamps, a core property of log entries and an important feature for forensic analysis.
In our scheme, we separate logs into epochs, e.g., hourly, and construct a hash tree per epoch that contains all log entries for the epoch as leaves.
Our hash tree construction also stores additional parity information divided into sub-epochs, e.g., on a sub-second basis, which allows to localize alterations to the logs at sub-epoch level, even though there is only one tree root per epoch.

This tree construction allows to identify tampering of a set of logs by only storing the tree's root and additional parity information, resulting in a constant storage overhead regardless of the number of logs.
Further, it allows constructing \acp{poi} that serve as \emph{receipts} for individual log entries, which can be stored by interacting nodes to later prove that an event has happened and was logged, even if the original set of logs was destroyed.

\begin{figure}[t]
	\centering
	\includegraphics[width=1\columnwidth]{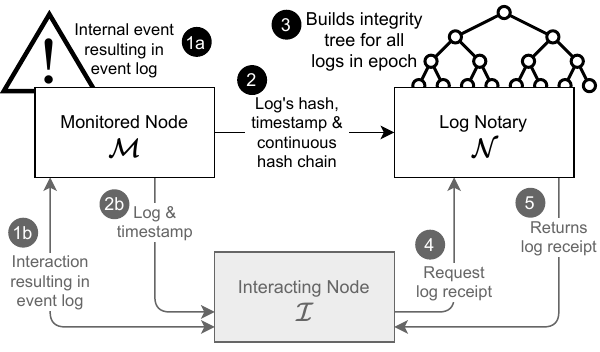}
	\caption{Overview of steps in our scheme for both internal and interaction events.}
	\label{fig:overview}
\end{figure}

\Cref{fig:overview} outlines how our scheme works.
Step 1a shows an internal event happening on or measured by the monitored node \monitored, which results in an event log entry.
In step 2, \monitored will submit the hash and timestamp of the log to the log notary \notary.
\monitored also sends a hash that is part of a continuous hash chain to ensure \emph{forward integrity}~\cite{bellare1997forward}.
This ensures that any tampering of logs between the last and the current submission will become apparent (e.g., if the adversary tries to alter logs previous to the compromise that have not yet been sent out).
Thus, if the forward integrity check fails, \notary will record the affected logs for a later investigation.
After the current epoch has ended, \notary will build this epoch's integrity tree with all logs received, construct the respective sub-epoch integrity information, and publish the tree's root in step 3.
After all interactions for an epoch were performed, \notary can discard the epoch tree, except for its root and parity information, resulting in a constant storage overhead.

In case of an interaction event with an interacting node \interacter---gray in \Cref{fig:overview}---the interaction will result in an event log entry in step 1b.
Besides sending the log to \notary in step 2, \monitored will also send the entire log to \interacter as well in step 2b.
After \notary builds the integrity tree, in step 4, \interacter will request a log receipt from \notary that is returned in step 5.
The log receipt gives \interacter proof that the log was indeed recorded by \notary.

\begin{figure}[t]
	\centering
	\includegraphics[width=1\columnwidth]{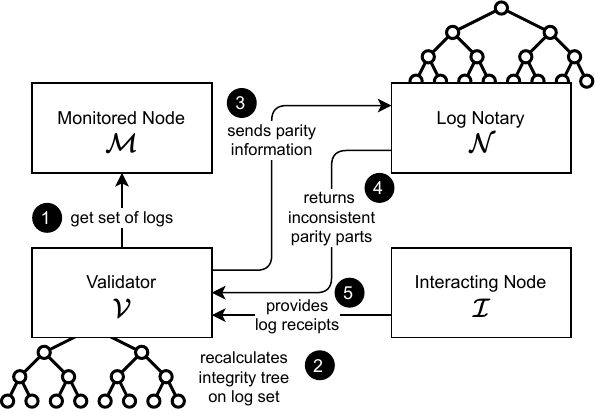}
	\caption{Overview of steps for validating logs.}
	\label{fig:overview_validator}
\end{figure}

At a later point in time, validator \validator wants to investigate an event that involves \monitored's logs, depicted in \Cref{fig:overview_validator}.
To identify any tampering, \validator first accesses the current---and potentially altered---set of logs on \monitored (step 1) and recalculates the respective epoch's tree on this set (step 2).
In case the tree's root is the same, \validator has proof that the set of logs on \monitored has not been tampered with.
If the re-calculated tree root differs, \validator calculates the sub-epoch integrity data for the current set of logs and sends it to \notary in step 3.
In step 4, \notary compares this information with its stored integrity data and returns the identified inconsistent parts.
If \notary recorded some inconsistencies regarding the forward integrity check, it will also return these to \validator.
This allows \validator to localize the sub-epochs of any alterations to the set of logs.
In case the log pertains to an interaction, \interacter can also provide its log receipt to \validator (step 5).
This can both prove that a relevant interaction event actually happened and may even help \validator increase the accuracy of any located tampering.

\subsection{Notation}
\Cref{tab:notation} gives an overview of our notation throughout the paper.

\begin{table}[h]
	\caption{Overview of our notation.}
	\begin{tabular}{llll}
		\multicolumn{2}{l}{\textbf{Entities}} 	& \multicolumn{2}{l}{\textbf{Logs}}       \\ \hline \\[-.95em]
		\monitored 	& monitored node			& \epoch & epoch                                \\[.2em]
		\interacter & interacting node			& \subepoch & sub-epoch                            \\[.2em]
		\notary 	& log notary				& \elog & an event log                         \\[.2em]
		\validator 	& validator					& \elogsepoch & \parbox{30mm}{all \elog that happened \\[-.3em] in \epoch}                           \\[.6em]
		\adv 		& adversary					& \elogts & timestamp of \elog                      \\[.2em]
		&                                   	& \elogreceipt & receipt of \elog                         \\[1.2em]
		\multicolumn{2}{l}{\textbf{Tree}}     & \multicolumn{2}{l}{\textbf{Parameters}} \\ \hline \\[-.95em]
		\treeepoch & \parbox{\widthof{\treename tree containing}}{\treename tree containing \\[-.3em] all \elog $\in$ \elogsepoch}    & \depthts & \parbox{30mm}{bit size of log \\[-.3em] timestamps in \treename tree} \\[.6em]
		\treeroot & root hash of \tree & \depthparity & \parbox{30mm}{tree depth of parity \\[-.3em] information} \\[.6em]
		\treeparity & \parbox{\widthof{parity information}}{parity information\\[-.3em] of \tree} & \sizeparity & size of each parity part \\[.2em]
		\treesecret & parity secret of \tree & \depthupdate & \parbox{30mm}{tree depth of \\[-.3em] receipt update}
	\end{tabular}
	\label{tab:notation}
\end{table}

\subsection{\treename Tree}
\label{sec:tree}
The \underline{P}arity \underline{I}ntegrity storing \underline{T}ime-\underline{S}parse tree (\treename) is the underlying integrity tree used in our scheme.
The \treename tree \tree is a binary hash tree specifically designed to store logs.
\monitored eventually submits all logs to \notary and once the epoch \epoch has ended, \notary will construct the \treename tree \treeepoch, containing all logs that happened in \epoch.

\subsubsection{Time-Sparsity.}
First, we introduce the time-sparsity property via timestamps, which works similarly to the \ac{smt}~\cite{smt}.
Instead of appending elements in a specific sequence to the tree---like with Merkle trees---the \ac{smt} constructs a complete tree containing all possible hash values of the respective hash function $h$, for instance, a tree with $2^{256}$ leaves for SHA256.
At first, all leaves are assigned to be empty, i.e., $h(\varnothing)$.
When a new element is inserted, the corresponding empty leaf is replaced with the element's hash and its position in the tree is defined by its hash.
While a tree of this size seems infeasible to compute, most leaves will be empty, i.e., the tree will only be \emph{sparsely} populated.
Thus, we can omit its empty parts, making it similarly efficient to compute to a traditional Merkle tree.

Instead of using the elements' hashes, the \treename tree uses timestamps to determine the position of logs' hashes in the tree.
Thus, the \emph{address} of the log entry is its timestamp and the \emph{content} of the log entry is its hash.
We need to define how many bits the timestamps need to represent logs with sufficient accuracy.
We define this parameter as \depthts.
For example, if our logs' timestamps \elogts have millisecond resolution, then we set \depthts to 22~bits to cover all possible timestamps within \epoch set to hours (3.6 million milliseconds per hour).
Similarly to \ac{smt}, we construct a hash tree that already addresses all possible timestamps as leaves and results in a tree with a height of \depthts.

\begin{figure}[t]
	\centering
	\includegraphics[width=1\columnwidth,trim={50 30 120 10},clip]{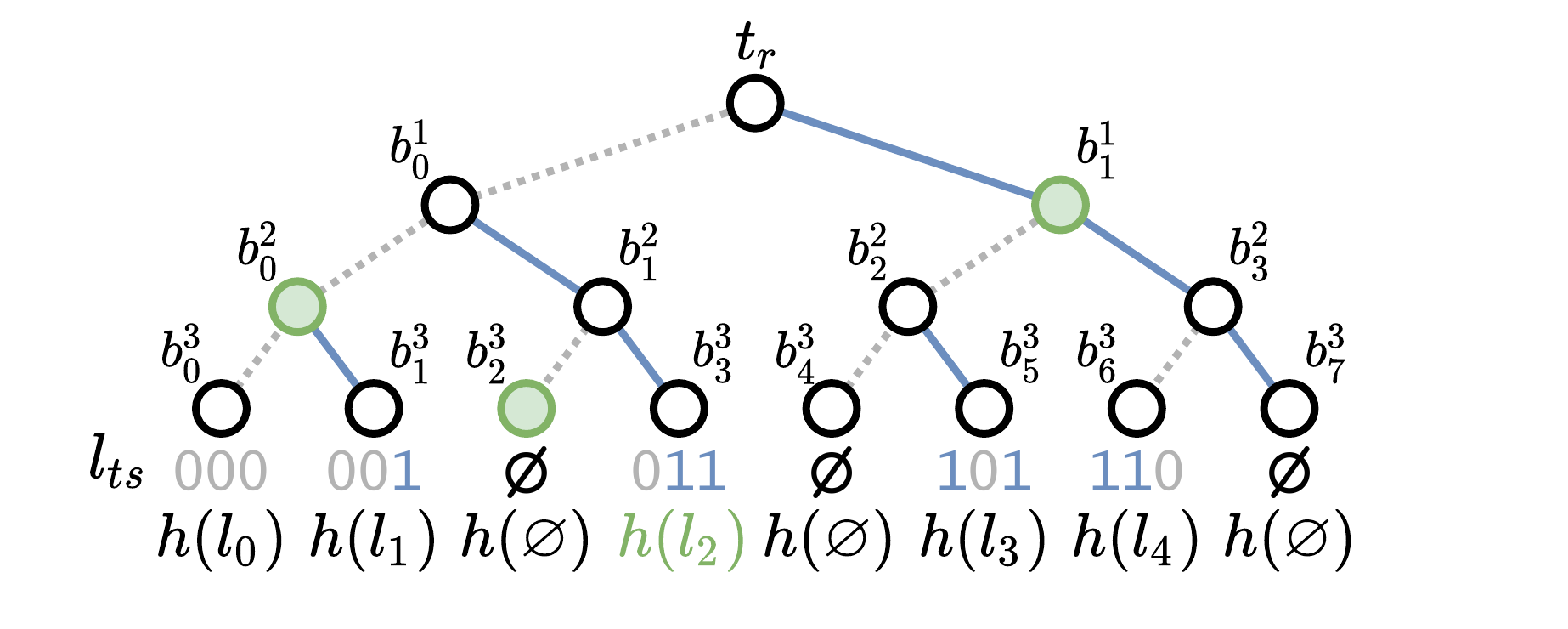}
	\caption{Exemplary \treename tree with \depthts = 3~bits and five events $l_0, \dotsm, l_4$.} 
	\label{fig:tree_structure}
\end{figure}

\Cref{fig:tree_structure} shows a simplified \treename tree as an example, containing logs with 3~bits timestamps.
This demonstrates how the deterministic positioning of the logs in the tree works.
To address specific nodes in the tree, we will use the notation $b_x^y$, where $y$ addresses the tree depth counting down, with the root being 0 and the leaves being \depthts, and $x$ addresses the respective $x$th hash on the $y$th depth level.
Take log entry $l_3$, for example, which happened at timestamp 5, or 101 in binary.
As the first bit is set, the path to the corresponding leaf from the tree's root \treeroot goes first right to $b_1^1$, along the solid blue line, then to the left to $b_2^2$ due to the next bit being unset, along the dotted gray line, and finally takes another right to $b_5^3$.
This leaf stores the hash $h(l_3)$ of the log.
Timestamps without any logs simply have $h(\varnothing)$ assigned.
When both leaf siblings are empty, the next depth of the tree gets $h(h(\varnothing)||h(\varnothing))$ assigned, and so on for the other tree depth levels.

\Cref{fig:tree_structure} also depicts how a log receipt \elogreceipt is constructed for $l_2$, i.e., the leaf $b_3^3$.
To be able to calculate \treeroot without knowledge of the entire tree, \elogreceipt contains the three hashes $[b_2^3, b_0^2, b_1^1]$, which are marked in green in the figure.
With knowledge of $l_2$ (including its hash) and its timestamp \elogts, \validator can construct the node $b_1^2$ with the help of $b_2^3$, and so on for the other hashes in the list until \treeroot is calculated.
Assuming \validator knows the actual \treeroot, e.g., via publication by \notary, it can compare it with the newly calculated root, and if they match \elogreceipt is valid, i.e., $l_2$ is proven to be part of the set of logs.
The number of hashes required to construct \elogreceipt is the same as the tree height, and thus the timestamp size \depthts, e.g., 22~bits or 22 hashes per \elogreceipt.
However, the leaves will only be \emph{sparsely} populated, i.e., many leaves will not have logs assigned to them; thus, empty parts of the tree can be omitted, e.g., $b_2^3$ in the figure, for the receipt of $l_2$.
To reflect these omissions, \elogreceipt also requires a bitmap for validation, e.g., 22~bits in size, to mark which tree depth levels were omitted due to the respective sibling branches being empty.
On average, each \elogreceipt requires log(\elogsepoch) hashes, and thus, fulfills requirement~\hyperref[req2:receipts]{R.2}.
In \Cref{sec:ops}, we give the pseudocode for the construction of the tree and log receipts.

\paragraph{Duplicate Timestamps.}
It might be the case that two or more logs share the same timestamp.
In this case, log entries may share a leaf in the tree, i.e., multiple log entries, ordered by their hashes, which can be concatenated into a single leaf hash.
Therefore, a \elogreceipt covering a log contained in such a leaf additionally requires the other hashes to be valid.
In anticipation of these cases, the \depthts parameter can be increased to also contain a counter.

\subsubsection{Parity Information.}
\label{sec:parity}

In our example, we set epoch \epoch to be hours, implying that there will be a new tree root \treeroot generated by \notary every hour.
Yet, in case \adv tampers with the original set of logs, an invalid \treeroot will merely reveal that some alteration was made sometime in that hour.
Even though this does not affect any \elogreceipt that \interacter holds, to increase the resolution of \validator's integrity check, we add additional parity information to the final tree.
For this, we set the size of the parity \sizeparity, defining how many bits per sub-epoch \subepoch are stored, and the parity depth \depthparity, defining at which tree depth---counting from the root down---the parity information will be calculated.
After the tree's root \treeroot and parity \treeparity are calculated, i.e. when the current \epoch ends, \notary can delete the rest of the tree, while still able to do integrity checks to identify any tampering with the logs.

\begin{figure}[t]
	\centering
	\includegraphics[width=1\columnwidth,trim={50 30 120 40},clip]{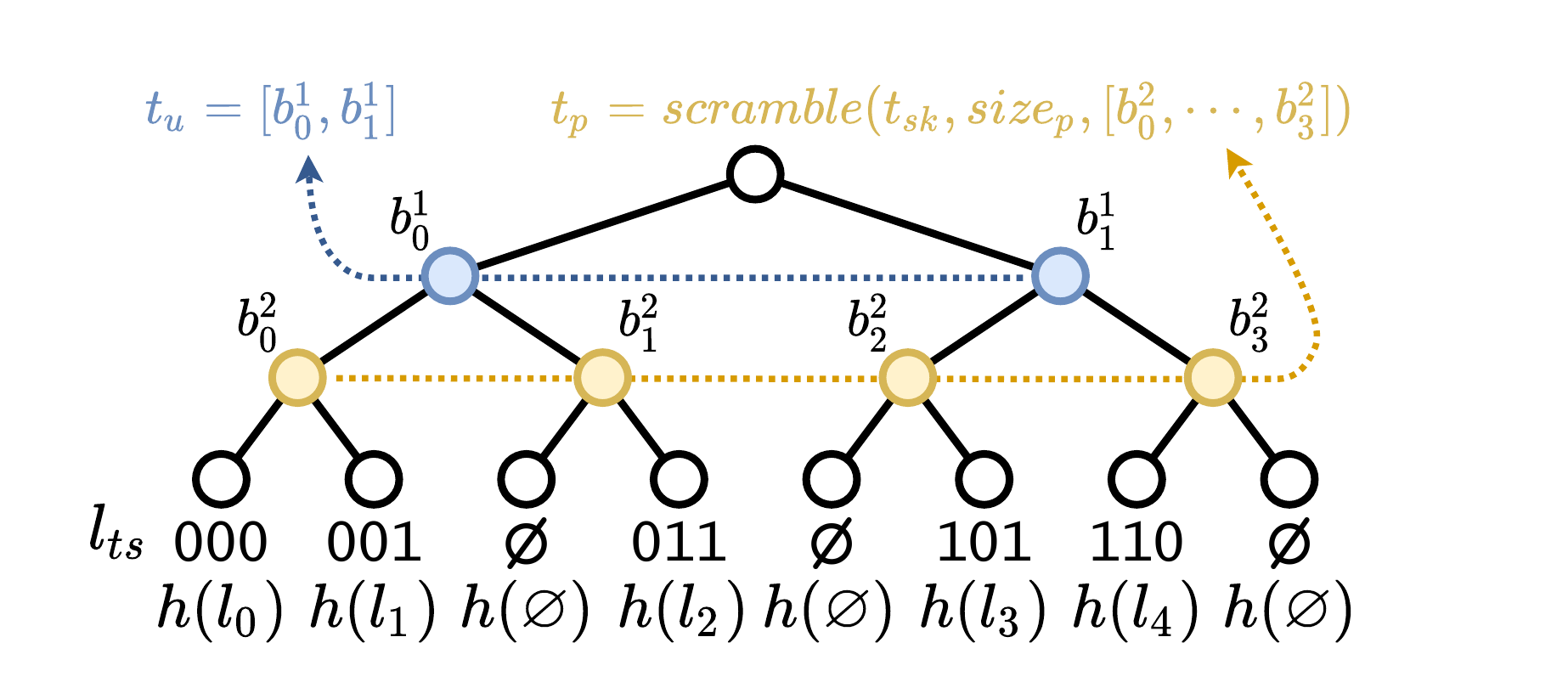}
	\caption{Construction of the parity \treeparity and universal update \treeupdate in the example tree as in \Cref{fig:tree_structure}.} 
	\label{fig:tree_parity}
\end{figure}

\Cref{fig:tree_parity} shows how the parity information is constructed in the \treename example tree.
In the case shown in the figure, \depthparity $=2$, and thus, we look at the four hash nodes $b_0^2, \dotsm, b_3^2$ at the second depth level of the tree (cf. yellow nodes in the figure).
In this case, \epoch is divided by four \subepoch.
In a naive approach, we could simply take the last \sizeparity bits of each of the \subepoch hashes and store these as the parity along with \treeroot.
If the original set was tampered with, e.g., a single log was deleted, we can locate the alteration by re-calculating the parity on the changed set of logs, comparing it to the original parity, and identifying the altered \subepoch.
For example, if $l_3$ was deleted, the hash of $b_2^2$ will be different, and thus, the corresponding parity will reveal an alteration in the third \subepoch.
In our example, with hourly \epoch and sub-second \subepoch, this would imply \depthparity$=12$ (4096 \subepoch) and if we set \sizeparity = 16~bits, would result in additional 8KB of parity stored per hour.
The additional parity allows \validator to detect any tampering with sub-second resolution.

However, this naive approach is insufficient as \adv can anticipate this check.
For example, \adv can simply calculate the parity on the original set---before any alterations---and calculate the expected parity for the targeted \subepoch.
\adv can then replace the deleted log with another log and randomly change the fraudulent log's data until \adv finds a collision resulting in the same parity.
While the \treeroot will still be invalid, \validator cannot accurately locate the alteration, as the parity will appear valid.
To prevent \adv from easily hiding changes from the high-resolution parity, we introduce the tree's parity secret key \treesecret for the tree's parity \treeparity.
Instead of simply taking the last bits of each parity hash, \treesecret contains bit positions that will be extracted from each \subepoch hash.
\treesecret is randomly generated by \notary and consists of \sizeparity numbers between 0 and the bit-length of the used hash function's digest, such that all bit positions can be addressed.
For our example, with SHA256 and 2~Bytes of parity per \subepoch, i.e., \sizeparity$=16$~bits, \treesecret would consist of 16 numbers from 0 to 255.
As \notary does not reveal \treesecret, \adv is unable to purposefully find a collision, excluding a 1-in-65536 chance of being lucky, in our example.
This allows our scheme to set a time resolution for the tamper detection via \epoch, and thus, fulfills requirement~\hyperref[req1:tamper]{R.1}.

\subsection{Interactions}
\label{sec:ops}
\begin{figure*}[p]
	\centering
	\includegraphics[height=\textheight,trim={0 0.15cm 0 0},clip]{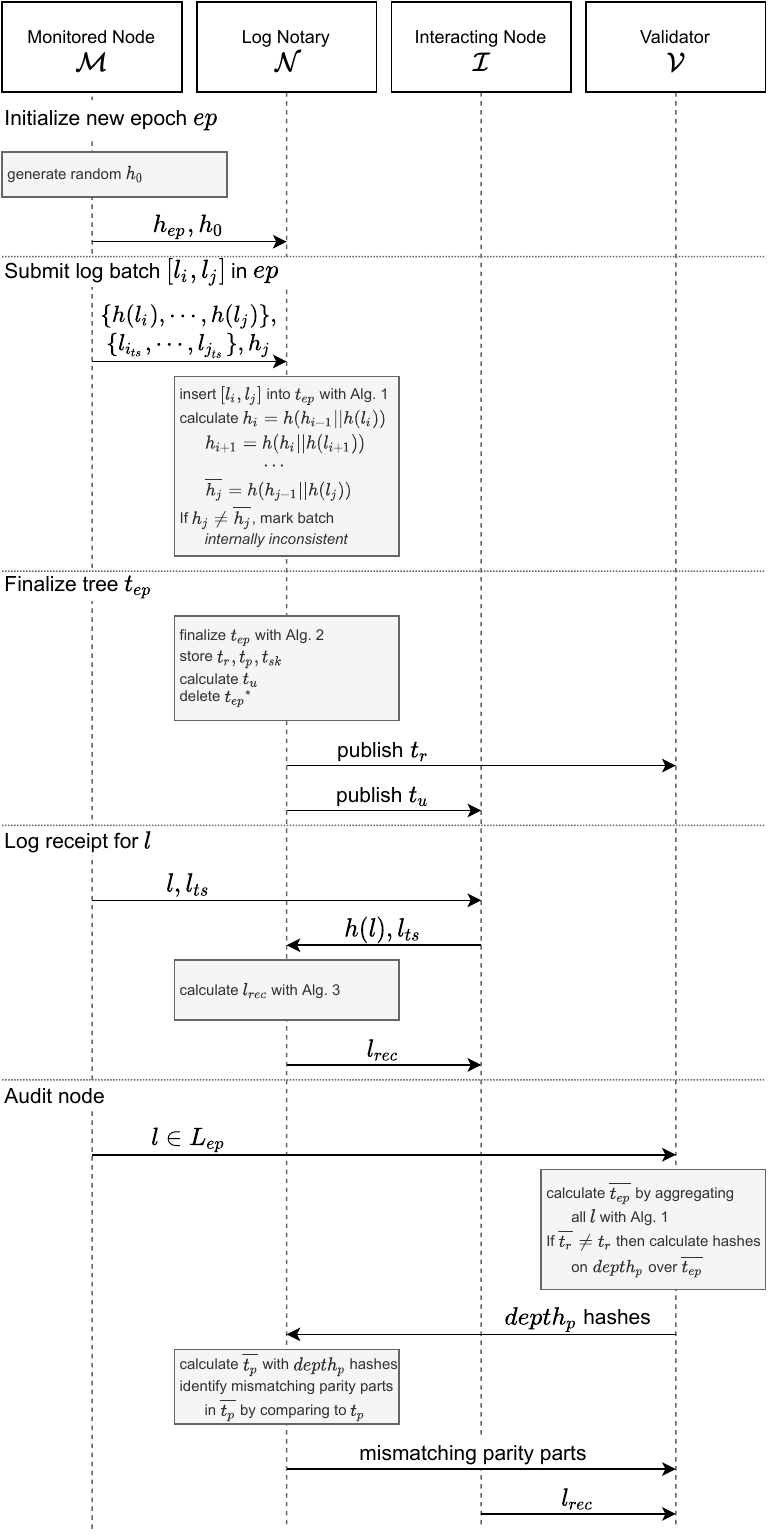}
	\caption{Overview of interactions in our scheme.} 
	\label{fig:protocol}
\end{figure*}

In this section, we define the interactions between the actors that constitute our tamper-evident logging scheme.
In the following, we describe what happens when \monitored submits a log to \notary, how a \treename tree is finalized by \notary after an epoch has ended, how \interacter gets a log receipt, and how a validator audits \monitored.
\Cref{fig:protocol} shows an overview of these interactions.

\subsubsection{Submit Log.}
\label{sec:ops:log}
This operation is executed between \monitored and \notary to submit a log \elog in \elogsepoch for the current epoch \epoch.
To initiate the operation, \monitored hashes \elog and sends \eloghash as well as the log's timestamp \elogts to \notary.
As only the hash of the log is sent, \notary learns nothing about the log, besides its timestamp, which ensures the privacy of \monitored.
\notary will then insert the log in the current \treeepoch, as shown in \Cref{alg:add_event}.
Conveniently, due to the sparse structure of the \treename tree, to address the correct node's position in the tree for the respective depth, we simply need to cut off $(size_{ts}-y)$ bits in $x$, with $b_x^y$.
For example, to address the correct node in the tree for \elogts on depth 8, we simply take the first 8~bits in \elogts as $x$ when addressing $b_x^8$.
To outline the algorithm, it will insert \eloghash at the respective leaf and progressively update the affected branches for each depth level of the tree.
For this, the respective bit in \elogts is flipped to obtain the corresponding sibling of the current node.
This process is repeated until \treeroot, i.e., $b_0^{size_{ts}}$, is reached.
When all logs in \elogsepoch were processed, \treeepoch will be complete and finalized in the next section.

\begin{algorithm}[t]
	\caption{\texttt{add\_log} function executed by \notary to add an event log \elog to the respective \treeepoch}
	\label{alg:add_event}
	\footnotesize
	\raggedright
	\begin{algorithmic}[1]
		\Require \eloghash, \elogts
		\Ensure \treeroot
		
		\State $b_{l_{ts}}^{size_{ts}}$ $\gets$ \eloghash
			\Comment{set the corresponding leaf to the log's hash}
		\For{$i \gets size_{ts}$ \textbf{to} $1$}
			\Comment{update all tree depth levels from the bottom up}
		\State \emph{sibling} $\gets$ \elogts with the $i$th bit flipped
		\If{$i$th bit set in \elogts}
			\Comment{get the respective neighbor hashes}
			\State \emph{left} $\gets$ $b_{sibling}^{i}$
			\State \emph{right} $\gets$ $b_{e_{ts}}^{i}$
		\Else
			\State \emph{left} $\gets$ $b_{e_{ts}}^{i}$
			\State \emph{right} $\gets$ $b_{sibling}^{i}$
		\EndIf
		\State $b_{e_{ts}}^{i-1}$ $\gets$ $h($\emph{left} $||$ \emph{right}$)$
			\Comment{compute the branch in the next depth}
		\EndFor
		
		\State \textbf{return} $b_0^{size_{ts}}$
	\end{algorithmic}
\end{algorithm}

\paragraph{Forward Integrity.}
Nevertheless, in a practical application, it may occur that \monitored is not always connected to \notary.
Thus, \monitored is likely to irregularly send batches of logs to \notary.
However, this implied delay between logging an event and submitting the respective batch of logs to \notary opens a vulnerability window.
\adv may compromise \monitored and alter logs that were not yet submitted to \notary, preventing \validator to identify any alteration.
To avoid this, we augment the log submission to ensure forward integrity.
To ensure this, at the start of \epoch, \monitored will select a random hash $h_0$ to start a hash chain\footnote{It can also just be a sufficiently large random number.}, send $h_0$ to \notary, create a log for starting a new hash chain $l_c$, and overwrite $h_0$ with $h_1=h(h_0 \,||\, h(l_c))$.
All subsequent logs $l_i$ will continue the hash chain and overwrite the old value $h_{i-1}$ with $h_i=h(h_{i-1} \,||\, h(l_i))$.
When \monitored sends a batch of logs to \notary, it will additionally send the current hash chain value $h_i$.
Knowing either $h_0$ or the last sent hash chain value, \notary can calculate the expected hash chain value by processing all log hashes contained in the batch.
If the resulting hash chain value matches the $h_i$ received from \monitored, \notary can confirm there were no alterations between log creation and submitting the batch.
Otherwise, \notary marks the batch as \emph{internally inconsistent} and stores it for any later forensic analysis by \validator.
This prevents \adv from tampering with any logs recorded prior to \monitored's compromise, as \adv cannot obtain the overwritten $h_{i-1}$ required to construct a valid $h_i$.

Additionally, on each new \epoch, \monitored will also send the last hash in the chain of the previous epoch $h_{ep}$, along with $h_0$.
This prevents \adv from truncating unsent logs on \monitored, and thus, convincing \notary that no more log entries were generated since the last received batch of logs.
Expecting a valid $h_{ep}$ prevents \adv to achieve this unnoticed.
As already-submitted logs are protected by \treeepoch, this mechanism is sufficient to ensure forward integrity.

\subsubsection{Finalize Tree.}
\label{sec:ops:finalize}
After \notary received all logs in \elogsepoch and the hash chain hashes closing \epoch, it will finalize the tree and calculate the parity for later subepoch \subepoch integrity checks.
\Cref{alg:finalize} shows how the parity calculation works.
As described in \Cref{sec:parity}, \notary first generates the tree's parity secret \treesecret by randomly generating \sizeparity numbers, able to address any bit in a hash digest.
In our example, we chose \sizeparity of 16~bits and SHA256, and thus \treesecret contains a list of 16 single Byte integers.
\notary then calculates the parity share for each hash at the tree depth \depthparity, by taking the bits addressed by \treesecret.
Thus, in our example with \depthparity at tree depth 12, \Cref{alg:finalize} calculates the 16 bit sized parity share for the hashes $b_0^{depth_p}, \dotsm, b_{4095}^{depth_p}$, resulting in \treeparity size of 8KB.
While \treeparity and \treesecret are not published by \notary, \treeroot can be published to allow \validator to validate any \elogreceipt without directly involving \notary.

\begin{algorithm}[t]
	\caption{\texttt{finalize\_tree} function executed by \notary to calculate \treeepoch's parity}
	\label{alg:finalize}
	\footnotesize
	\raggedright
	\begin{algorithmic}[1]
		\Require \treeepoch
		\Ensure \treeroot, \treeparity, \treesecret
		
		\State \treesecret $\gets$ empty list of integers
		\State \treeparity $\gets$ empty list of raw bits of length \sizeparity each
		\For{$i \gets 0$ \textbf{to} \sizeparity}
			\Comment{generate secret}
			\State \emph{rnd} $\gets$ unique random number between 0 and bitsize of hash digest
			\State \treesecret\texttt{.append(}\emph{rnd}\texttt{)}
		\EndFor
		
		\For{$i \gets 0$ \textbf{to} $2^{depth_p}$}
			\Comment{construct parity}
			\State \emph{par} $\gets$ 0
			\For{$rnd$ \textbf{in} \treesecret}
				\State \emph{par} $\gets$ \emph{par} $|$ (1 $\ll$ ($rnd$th bit in $b_i^{depth_p}$))
					\Comment{copy respective bit}
			\EndFor
			\State \treeparity\texttt{.append(}\emph{par}\texttt{)}
		\EndFor
		
		\State \textbf{return} $b_0^{size_{ts}}$, \treeparity, \treesecret
	\end{algorithmic}
\end{algorithm}

\subsubsection{Log Receipt.}
\label{sec:ops:receipt}
If any log \elog recorded by \monitored is of interest to \interacter, it will request the receipt \elogreceipt from \notary.
For this, \interacter will send \notary the hash of the log \eloghash and \elogts, not revealing anything about the nature of the underlying event to \notary.
Calculating \elogreceipt is shown in \Cref{alg:calc_poi}.
\notary will progressively check the sibling hashes along the path of \elogts.
If a non-empty sibling was found, the sibling hash is added to the proof of inclusion hash list.
If a sibling is empty, it is simply skipped.
Nevertheless, for \validator to be able to correctly reconstruct the path, it also needs to know on which tree depth levels siblings were skipped, i.e., empty.
Thus, \Cref{alg:calc_poi} also builds a bitmap of size \depthts, e.g., 22~bits for millisecond logs.
Every time a sibling is added to the proof of inclusion, the respective bit for the current tree depth is set.
\notary will then send \interacter both the proof list and the bitmap, i.e., \elogreceipt.

\begin{algorithm}[t]
	\caption{\texttt{calc\_receipt} function for calculating the receipt of a log}
	\label{alg:calc_poi}
	\footnotesize
	\raggedright
	\begin{algorithmic}[1]
		\Require \eloghash, \elogts
		\Ensure \emph{poi}, \emph{poi\_bitmap}
		
		\State \emph{poi} $\gets$ empty list
		\State \emph{poi\_bitmap} $\gets$ 0
		\For{$i \gets 0$ \textbf{to} \depthts}
			\Comment{store non-empty hashes for all tree depth levels}
			\State \emph{sibling} $\gets$ \elogts with the $i$th bit flipped
			\If{$b_{sibling}^{i}$ $\neq$ $\varnothing$}
			\Comment{$\varnothing$ is the empty hash for the respective tree depth}
				\State \emph{poi}\texttt{.append(}$b_{sibling}^{i}$\texttt{)}
				\State \emph{poi\_bitmap} $\gets$ \emph{poi\_bitmap} $|$ (1 $\ll$ i)
			\EndIf		
		\EndFor
		
		\State \textbf{return} \emph{poi}, \emph{poi\_bitmap}
	\end{algorithmic}
\end{algorithm}

\paragraph{Receipt Provision.}
Receiving the log receipt \elogreceipt requires a complete \treeepoch, and thus \elogreceipt can only be calculated after the tree was finalized, i.e., the current \epoch has ended.
However, it may be impractical to wait for the current epoch to end until receiving \elogreceipt, e.g., \elog happened at the beginning of the epoch and \interacter does not want to wait an hour for \elogreceipt to finish a transaction depending on it.
A simple solution is to rely on a commitment from \notary that \eloghash was submitted and a promise that \elogreceipt will be sent when the epoch has ended.
While this may suffice, we additionally introduce the concept of a generic \emph{receipt update} to improve efficiency.
Instead of waiting for the epoch \epoch to end, we introduce the parameter \depthupdate that, similarly to \depthparity, defines the tree depth for the receipt update.
Due to the timestamp-based nature of \treename, this implies that \epoch is divided into $2^{depth_u}$ parts.
When \interacter requests \elogreceipt of a log that has just occurred, it merely has to wait until the current branch on depth \depthupdate is final, and then receives a partial \elogreceipt up to tree depth \depthupdate.
In our example, when \depthupdate is set to 10 it divides the tree into 1024 parts, and thus, a new branch is started every $\sim$3.5s.
The partial \elogreceipt is then potentially missing 10 tree depth levels to calculate \treeroot.
After the current epoch has ended, \notary will publish the receipt update \treeupdate, containing all hashes at tree depth \depthupdate, as shown in \Cref{fig:tree_parity} for the simple \treename tree.
In our example, 1024 hashes with 256~bits each result in a \treeupdate of 32KB.
\treeupdate can also be provided by any untrusted third party, as \treeupdate can be validated by checking if it results in \treeroot, which is published by \notary.
This allows all \interacter to finalize \elogreceipt by themselves, calculating the missing siblings to \treeroot with the help of \treeupdate.
The finalized \elogreceipt can be validated against \treeroot.
This avoids any specific commitment scheme or many individual requests to \notary after an epoch has ended while ensuring privacy.

\subsubsection{Audit Node.}
\label{sec:ops:audit}
When \validator inspects \monitored to find any alterations in a given epoch \epoch, it needs to extract all logs in \elogsepoch on \monitored.
\validator then calculates \treeepoch from this set, as in \Cref{alg:add_event}.
If the recalculated \treeroot matches the \treeroot published by \notary, the logs on \monitored were not tampered with.
If they do not match, \validator calculates all hashes on tree depth \depthparity and sends it to \notary.
\notary will then recalculate the \treeparity of this altered set by executing \Cref{alg:finalize} on it with the stored \treesecret.
This allows \notary to compare all parity shares in both \treeparity and any mismatches are reported back to \validator, such that it can localize the specific \subepoch in which the alteration has happened.
Additionally, \validator may contact \interacter and request any \elogreceipt to confirm individual events.
If any acquired \elogreceipt have a \elogts close to the alteration, it can additionally help to further pinpoint when the alteration happened.
For example, if a \elogts is mostly along the path of a, e.g., deleted log, a hash deviation on a higher depth than \depthparity can indicate a more accurate time of the alteration.


\section{Security}
\label{sec:security}
The goal of adversary \adv is to tamper with the event logs of the monitored node \monitored to cover up an event incriminating \adv.
Further, \adv tries to circumvent the accurate localization of this tampering by the validator \validator to hamper any efforts to collect indirect evidence, which could lead to raising suspicion towards \adv.
To achieve this, \adv can employ four different strategies.
\adv may try to (1) retroactively cover up the tampering, (2) hide the accurate time of the alteration, (3) tamper with the logs before they are sent to \notary, or (4) obfuscate the alteration by strategically performing many alterations.

To retroactively cover up the tampering without raising any suspicion (1), \adv has to tamper with \monitored's logs in a way such that the altered set of logs will result in the same \treeroot as published by \notary.
However, we assume Merkle tree constructions are collision-resistant (cf. \Cref{sec:advmodel}); thus, this strategy will fail.

Nevertheless, \adv may accept that \treeroot shows that tampering has happened, yet, \adv may try to circumvent the parity check via \treeparity to prevent \validator from finding the sub-epoch of the tampering (2).
To achieve this, \adv has to tamper with the set of logs on \monitored, such that the recalculation of \treeparity by \notary will result in the same parity information.
However, as \adv does not have access to \treesecret or \treeparity, \adv cannot purposefully reproduce a set of logs resulting in the same \treeparity.
As explained in \Cref{sec:parity}, there is merely a 1-in-$2^{size_p}$ chance that \adv may get lucky and get a matching \treeparity, e.g., if \sizeparity = 16~bits, a 1-in-65536 chance.
This chance applies for a single altered log entry and any subsequent alteration reduces this chance exponentially.

Furthermore, \adv may prevent \monitored to send the incriminating log in the first place (3), by either tampering with a log batch before it is sent or by simply blocking \monitored's communication until the alteration is accomplished.
Yet, the forward integrity measure, as described in \Cref{sec:ops}, will prevent this from succeeding.
As \monitored will overwrite any past hashes in the hash chain, \adv is unable to alter the unsent log batch and provide a valid hash to convince \notary that no logs were altered between the last submitted batch and the submission of the altered batch.
\adv may try to execute a \emph{truncation attack}~\cite{ma2009new} by suppressing an unsent log batch and all further log submissions for the current epoch, essentially trying to convince \notary that the latest received log batch and forward integrity hash were the last for the current epoch.
However, as described in \Cref{sec:ops:log}, on closing an epoch, \notary expects a valid chain hash, which \adv cannot generate for any logs before the compromise.
Thus, the truncation will not go unnoticed by \notary.
\adv will not be able to intercept any hash of the forward integrity hash chain (i.e., $h_0$ or any $h_i$) before compromising \monitored, as their communication is protected (cf. \Cref{sec:advmodel}).

Finally, \adv may try to hide an alteration to the logs by introducing multiple alterations obfuscating the incriminating alteration (4).
Thus, \adv may change logs in many---or even all---sub-epochs such that \treeparity will indicate many altered sub-epochs, exacerbating \validator's efforts to localize the incriminating time of a specific alteration.
Additionally, \adv could also simply delete all logs.
Similar to other works in this area, our scheme cannot completely protect against this strategy; yet, it prevents \adv from \emph{stealthily} covering up the alteration and forces \adv to conspicuously tamper with \monitored.
Further, log receipts by any interacting node \interacter are still valid, even if all logs are deleted, and may help \validator to collect evidence about an incident.

\section{Evaluation}
\label{sec:eval}
In this section, we will evaluate the performance of our scheme.
First, we will discuss the performance overheads of the monitored node \monitored.
Afterwards, we will focus on the performance overhead of the log notary \notary in terms of computation, storage, and memory.
Finally, we will briefly discuss the performance impact of other parameters.

\subsection{Overheads on Monitored Node}
\label{sec:eval:monitored}
Besides the overhead of the actual logging process, \monitored needs to compute two hashes per log entry submission, \eloghash and the next hash in the hash chain ensuring forward integrity, as described in \Cref{sec:ops:log}.
For network overhead, \monitored sends a hash per log entry to \notary and a hash chain hash per batch, i.e., when multiple log hashes are sent in bulk.
The only additional storage overhead on \monitored is the current hash in the forward integrity hash chain.
The beginning of a new epoch requires \monitored to generate and send a random hash to \notary (see $h_0$ in \Cref{sec:ops:log}).
Overall, this shows that our scheme fulfills requirement~\hyperref[req3:overhead]{R.3}.

\subsection{Computational Overhead.}
\label{sec:eval:computation}
To analyze the computational overhead of \notary, we implemented a prototype of the \treename tree in Python.
We then executed scalability tests for the individual operations (cf. \Cref{sec:ops}) on a server with an AMD EPYC 7742 64-core CPU @~2.25~GHz and 256~GB RAM.
In the following, we present our measurements suggesting how many \monitored devices a single \notary server can handle in terms of computational and network overhead.
For all of our tests, we take the same parameters as our example in \Cref{sec:desgin}.
Thus, we take SHA256 as the hash function, \depthts~=~22~bits, \depthupdate~=~10, \depthparity~=~12, and \sizeparity~=~16~bits.

First, we measured how many logs our server can process per second, i.e., include the log's hash into a \treename tree with its time\-stamp.
For this test, we let 128 Python threads process as many logs as possible for 10 minutes, after which our results converged without significant deviations.
Note that after a thread processed one million logs, it will start a new \treename tree and continue.
We found that our server is overall able to process $718\,703$ logs per second on average and processing an individual log took 0.181ms on average.

Second, we measured the overhead for the \treename tree finalization, i.e., calculating the tree's parity information, as the tree root calculation is already covered by the processing of logs.
To measure this, for each run on a single thread, we generated a \treename tree with one million logs, measured the time it takes to finalize the tree, and repeated this process $100\,000$ times.
Our results show that a tree finalization takes 9.787~ms on average, or $\sim$78~s to finalize one million trees in parallel with 128 threads.

Finally, we measured the overhead of generating log receipts.
Here, we generate a \treename tree with one million logs and let 128 threads construct as many log receipts as possible for 10 minutes.
If the server has the entire \treename tree structure in memory, the server is able to generate $4\,484\,927$ log receipts per second.
We discuss the memory aspect further in \Cref{sec:eval:memory}.

\subsection{Storage Overhead.}
\label{sec:eval:storage}
The storage overhead of \notary is highly dependent on the chosen parameters.
If we note $|h|$ as the size of the hash digest, the storage overhead per epoch is given by:
\begin{alignat*}{3}
	t_r &{}+ t_p &{}+{}& t_{sk} &= \\
	|h| &{}+ (size_p \cdot 2^{depth_p}) &{}+{}& (size_p \cdot log_2(|h|)) &
\end{alignat*}

Note that \notary will delete the inner structure of \treeepoch eventually, and thus, only the tree's root hash and parity information needs to be stored.
We discuss this aspect further in \Cref{sec:eval:memory}.
In our example (cf. \Cref{sec:eval:computation}) this means a constant $8\,240$~Bytes per epoch, regardless of the number of logs.
If we further assume hourly epochs, the storage cost of a monitored node is $\sim$69~MB per year.
Thus, if we aim to cover one million monitored devices, \notary would need $\sim$66~TB of storage per year.
While the average size per log varies greatly for different use cases\footnote{We found numbers for average log sizes of 200~Bytes~\cite{logsize1}, 800~Bytes~\cite{logsize2}, 1024~Bytes~\cite{logsize3}, and 1500~Bytes~\cite{logsize4}.}, even if we assume a small average of 200~Bytes per log entry and $10\,000$ logs per epoch, our scheme reduces the storage overhead to less than 1\%, compared to storing the logs in their entirety.

The chosen size of the parity \sizeparity and the depth of the parity information \depthparity has the biggest influence on the resulting size.
With the same example and \sizeparity = 10~bits, these numbers are reduced to $5\,162$~Bytes per epoch and $\sim$43~MB per year.
However, this implies a security trade-off, as \adv has an increased chance of 1-in-1024 to find a matching \treeparity per altered log entry (cf. \Cref{sec:security}).
If we only alter \depthparity = 11, we get $4\,144$~Bytes per epoch and $\sim$35~MB per year; yet, the trade-off here is halving the time resolution of detecting alterations.

For longer-term storage, some simple optimizations can be made depending on the use case.
For example, we may assume that for logs older than a year forensic analysis will not be required anymore.
Thus, \notary could simply prune all parity information and only keep the tree roots for log receipt validation and per epoch tamper detection, e.g., reducing the required yearly storage to $\sim$274~KB per node or $\sim$261~GB for a million nodes.

\subsection{Memory Overhead.}
\label{sec:eval:memory}
While \notary will delete the inner structure of a \treename tree eventually, it should keep the entire tree for some time to be able to efficiently construct any log receipts \elogreceipt requested by an interacting node \interacter.
Alternatively, \notary may only keep the tree's leaves to save on memory and construct the tree's branch nodes each time a log receipt is requested.
As shown in \Cref{sec:eval:computation}, the latter can be efficient, if \notary does not expect a large amount of log receipt requests at a time.
For real-world deployment, it is also important to consider how long \interacter shall be able to request receipts, i.e., how long \notary will keep a tree before deletion.

\begin{figure}[t]
	\centering
	\includegraphics[width=1\columnwidth,trim={0 0 0 0},clip]{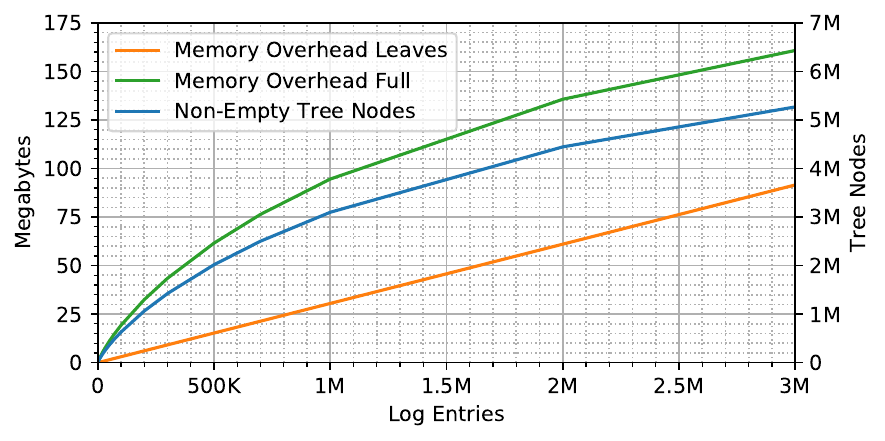}
	\caption{Graph showing the number of non-empty tree nodes (i.e., branch hashes) in a \treename tree with a number of log entries (i.e., leaf hashes). For the memory overhead, the graph shows storing the full tree with branches, as opposed to only storing the leaves.} 
	\label{fig:memory}
\end{figure}

\Cref{fig:memory} shows how much memory a single \treename tree may occupy, i.e., covering the logs for one epoch on a monitored node.
Note that we use the term \emph{memory} loosely in this context, as the ordered nature of the tree's leaves and branch nodes (i.e., ordered by timestamps) allows this data to be easily indexed and swapped to disk.
Based on measurements of our prototype, the graph shows the number of non-empty branch nodes in the tree as well as the memory overhead for both storing all nodes and only the leaves.
Considering these numbers, the memory overhead of \notary is significant.
For example, assuming each tree contains one million log entries, our server with 256~GB RAM can cover around 2700 full trees (94.58~MB per tree) or around 8500 trees with only leaves (30.52~MB per tree), without swapping to disk.
Note that, as we discuss in \Cref{sec:rw}, other works require storing per-log information, while our scheme requires a similar overhead only temporarily and eventually reduces the long-term storage overhead to the numbers shown in \Cref{sec:eval:storage}.

\subsection{Other Parameters}
\label{sec:eval:parameters}
In this section, we briefly discuss the performance impact of the other parameters in our scheme.
One is parameter \depthupdate, which is the depth of the receipt update, as described in \Cref{sec:ops:receipt}.
A bigger \depthupdate lowers the maximum waiting time of \interacter before \notary is able to construct a preliminary \elogreceipt.
Yet, each increment of \depthupdate also doubles the size of the update, and thus, the network overhead.
The size of the used timestamps \depthts for the \treename tree mainly affects the network overhead, as each a \elogts is sent along with each log entry (cf. \Cref{sec:eval:monitored}).
Further, it may also increase the number of non-empty tree branches in the \treename tree, which increases memory overhead if the entire tree is cached, as discussed in \Cref{sec:eval:memory}.


\section{Extensions}
\label{sec:extensions}
In this section, we discuss optimizations for practical deployments of our scheme.
\\~\\
\noindent\textbf{Omit Unused Parities.}
While we implied a constant influx of logs over time so far, many real-world applications experience times with bursts of logs and other times logging nothing at all.
This fact can be leveraged to improve storage efficiency by adding a parity bitmap, which marks empty sub-epochs.
This way the size of \treeparity can be easily reduced.
For example, assuming our monitored devices typically do not log anything at night, e.g., 8 hours from 10pm to 6am, the yearly storage reduces to $\sim$46~MB per device or $\sim$44~TB for a million devices, opposed to $\sim$66~TB if all sub-epochs are populated.
\\~\\
\noindent\textbf{Error Correction Code Parities.}
As we show in \Cref{sec:eval:computation}, the main influence on storage overhead is the parity information to be able to identify sub-epochs.
With the use of error correction codes (ECC)~\cite{ecc}, we found an alternative to construct the parity information that can reduce its size by reducing the number of altered sub-epochs we can detect.
Essentially, the idea is to treat alterations on \depthparity as transmission errors.
For example, to deploy a Reed-Solomon ECC~\cite{reedsolomon}, we first interleave individual Bytes on the \depthparity depth with a randomly chosen permutation.
On one hand, this transforms any alteration, e.g., altered hash of 32~Bytes, from a large burst error and spreads it among the entire \depthparity depth.
On the other hand, the used permutation can be used as the \treesecret, such that \adv cannot anticipate the parity information.
We treat the supplementary information added by the ECC as \treeparity stored by the log notary.
An altered parity set can then be \emph{error-corrected} and any corrected sub-epochs identified as alterations.
However, this approach limits how many altered hashes can be found, as too many consecutive errors will result in the ECC to fail.
A preliminary implementation showed that a (32,8) Reed-Solomon can find $\sim$150 altered hashes with high probability, while requiring $\sim$2.7~KB of parity information per tree, as opposed to $\sim$8~KB with the same parameters with our approach.


\section{Related Work}
\label{sec:rw}
A diverse set of approaches has been proposed in the area of tamper-resistance for logs, such as sending all logs to a remote service~\cite{rwmRemote2,rwmRemote3}, involving blockchains, or leveraging specialized hardware.
Remote logging services cannot detect attacks, in which logs are suppressed before they are sent to the remote service, e.g., by selectively blocking communication of the monitored device after compromise.
Blockchain-based approaches either store integrity proofs on slow and expensive permissionless blockchains~\cite{sutton2017blockchain,cucurull2016distributed}, use permissioned blockchains (i.e., a fixed set of validators) that essentially induce redundancy overheads~\cite{ahmad2018towards,ahmad2019blocktrail}, or hybrid approaches using both types of blockchains~\cite{aniello2017prototype,pourmajidi2018logchain}.
For our purposes, however, we focus on \emph{data structures} enabling tamper-evident logging.

\paragraph{Specialized Hardware}
One approach is to simply log to a Write Once Read Multiple (WORM) storage device~\cite{HWworm}.
In recent works, logging is secured by leveraging a local Trusted Platform Module 2.0~\cite{HWtpm} or using Intel's Trusted Execution Environment (TEE)~\cite{HWsgx}.
More recently, CUSTOS~\cite{HWcustos} generalizes TEEs for logging and additionally enables third-party verifiability as well as increase log availability.
In HardLog~\cite{HWhardlog}, a logging scheme leveraging an external logging device is used, which is protected against remote adversaries.
However, as we eluded to in requirement~\hyperref[req3:overhead]{R.3}, devices may be heterogeneous, and thus, requiring specific or external hardware is infeasible in many settings.
Note that tamper-evident data structures still provide value when combined with specialized hardware.
The schemes based on such data structures involve a trusted entity that can be replaced by trusted hardware, which has been considered~\cite{chong2003secure}.

\paragraph{Tamper-Evident Data Structures}
Bellare and Yee~\cite{bellare1997forward} first coined the term \emph{forward integrity} of audit logs and proposed a data structure that ensures this property.
Ensuring forward integrity means that an adversary is not able to alter log entries on a device that were committed before the compromise without being detected.
For example, the compromise itself may result in an access log, revealing some information about the adversary, and thus, such logs become an obvious target to cover up the compromise.
In the presented scheme, the logging device first generates a secret that is shared with a trusted verifier.
The logging device aggregates all logs in an epoch with a Message Authentication Code (MAC) using the shared secret and stores this MAC as well as an ongoing counter of log entries locally.
Afterward, the pre-shared secret is evolved in a deterministic way, such that the trusted verifier can replicate the resulting list of secrets for each epoch.
The new secret overwrites the previous one on the logging device and the entire process is repeated for each epoch.
This way an adversary is unable to produce a valid MAC after altering epochs that were committed before the compromise, while the verifier can verify the epoch MACs on the logging device due to its knowledge of the starting secret.

Schneier and Kelsey~\cite{schneier1998cryptographic,schneier1999secure} presented a similar approach using per-log MACs and a hash chain.
The logging device shares a secret with a trusted entity.
Each log entry is hashed, including the previous log's hash to form a chain, and these hashes are stored.
For every log hash, a MAC is calculated with the pre-shared secret, the resulting MAC is stored, and the secret evolves (overwriting the previous one).
Later, the MAC-based schemes were extended to use traditional public cryptography~\cite{holt2005logcrypt} to avoid the need to involve the trusted entity for validation.

However, the above schemes induce significant storage overhead on the monitored device, as the validation information stays on the device, violating requirement~\hyperref[req3:overhead]{R.3}.
Further, they were not designed with log receipts in mind.
While hash chains can provide this functionality in principle, they require sending the entire chain of logs (or rather their MACs or hashes) leading to the specific log to prove it is included.
Thus, proving the inclusion of individual logs is inefficient.
If the set of logs is internally consistent, skip lists can be used to achieve logarithmic log receipts~\cite{maniatis2002secure}.
However, to check internal consistency, the entire set of logs still needs to be inspected, as the skip links further along the list of logs may have been altered.
Thus, requirement~\hyperref[req2:receipts]{R.2} is not met with these schemes.
Nevertheless, these schemes were shown to be susceptible to \emph{truncation attacks}~\cite{ma2009new}, i.e., logs are only forward secure to the point they were committed to the trusted entity.
Thus, these schemes would need to frequently upload fresh proofs to the trusted entity to give the promised properties, which significantly increases network overhead compared to what is originally stated.
In contrast, our scheme has logarithmic-sized log receipts, small storage overhead on the monitored device (i.e., a single hash), and only requires a small constant amount of storage overhead on the trusted entity, as opposed to storage overhead on the monitored device linearly growing with the number of log entries.

The paper introducing the truncation attack also proposes an alternative approach leveraging a forward-secure sequential aggregate signature scheme~\cite{ma2009new} to achieve forward integrity without being susceptible to truncation attacks.
The approach allows the authentication of any log entry in the epoch with the resulting aggregated signature.
This requires only storing two MACs per epoch on the monitored device, and thus, greatly reduces storage overhead over previous works.
As this does not allow for efficient log receipts, the work also suggests an alternative mode of operation enabling efficient log receipts.
This mode requires additionally storing two MACs per log entry.
Thus, the two modes violate requirements~\hyperref[req2:receipts]{R.2} or \hyperref[req3:overhead]{R.3} respectively.
A similar approach based on Log Forward-secure and Append-only Signatures (LogFAS)~\cite{yavuz2012efficient} proposes a scheme with minimal computational and storage overhead on the trusted entity.
However, both approaches have been shown to be insecure (secret leakage for the former and signature forgery for the latter)~\cite{hartung2017attacks}.
Contrarily, our scheme only requires the monitored device to store a single hash and enables efficient generation of log receipts.

The work on the blind-aggregate-forward (BAF)~\cite{yavuz2009baf} logging scheme presents a forward-aggregation scheme based on elliptic curve public cryptography, which achieves efficient signature creation while not being susceptible to truncation attacks.
However, the scheme induces per-entry storage overhead on the trusted entity, requires large computational overheads for verification, and does not support fine-grained audits of individual logs or subset of logs; thus, it violates requirement~\hyperref[req2:receipts]{R.2}.
Our scheme supports log receipts, uses a configurable resolution on tamper-detection only affecting storage overhead on the trusted entity, which is constant with the number of log entries.

Another approach focuses on the efficient construction of log receipts with the \emph{history tree}~\cite{crosby2009efficient}, which is a Merkle tree with additional incremental proofs to prove consistency between two states of the tree over time.
The assumption is that all events involve clients that have interactions with the monitored device. Each interaction generates a log and the respective client always requests a receipt of this log.
There is no single trusted entity in the system, as many auditors collect log receipts; however, there has to be at least one honest auditor who sees \emph{all} log receipts.
To ensure all interaction logs are included, all clients need to immediately gossip each received log receipt, which allows the auditors to construct valid incremental proofs to ensure the history tree is consistent and complete.
Hence, this approach requires a significant communication overhead across the network as well as a lot of storage overhead on the monitored device, as it has to maintain the growing history tree.
The latter becomes also apparent in the work's evaluation~\cite{crosby2009efficient} as the required storage of the monitored device roughly doubles, besides the logs themselves.
Another work extends this approach to support non-membership proofs by introducing a \emph{hash treap} along with the history tree~\cite{pulls2015balloon}; yet, this approach shares the same downsides as the original approach~\cite{crosby2009efficient}.
Thus, these schemes violate requirement~\hyperref[req3:overhead]{R.3}.
On the contrary, the constant storage overhead of the \treename tree allows the trusted entity to scale to a large number of devices, as opposed to a large storage overhead on the monitored device.
Additionally, network overhead in our scheme is limited to one hash per log from the monitored device to the trusted entity, as opposed to many hashes per log to many entities in the network.


\section{Conclusion}
\label{sec:conclusion}
In this work, we presented a novel scheme for practical large-scale tamper-evident logging by proposing a new binary hash tree construction designed around timestamps.
This results in a constant storage overhead on a trusted third party with regard to a specified temporal resolution and enables the efficient construction of log receipts, which allow devices to keep proof of another device's event log.
We evaluated our scheme and showed that it requires significantly less overhead compared to previous works, allowing it to scale to a large number of devices.

\section*{Acknowledgment}
This work was supported by the European Space Operations Centre with Networking/Partnering Initiative and Huawei within the Resilient Event Integrity Assurance project.

\bibliographystyle{ACM-Reference-Format}
\bibliography{bib}

\end{document}